\documentclass[journal]{IEEEtran}
\usepackage{amsfonts}
\usepackage{verbatim}
\usepackage{mathrsfs}
\usepackage{amssymb,amsmath}
\usepackage[noblocks]{authblk}
\usepackage{stfloats}
\usepackage{algorithm}
\usepackage{algorithmic}
\usepackage{cite,url,subfigure,graphicx,wrapfig,verbatim,amsmath,amsfonts,amssymb,color, epsfig,graphics}
\usepackage{theorem}
\usepackage{array,color}
\usepackage{multirow}

\theoremheaderfont{\normalfont\bfseries}

\newcommand{\eg}{e.g.}

\ifodd 1
\definecolor{darkgreen}{RGB}{0,200,0}
\else

\fi

\begin{document}
\title{Resources on the Move for Smart City: A Disruptive Perspective on the Grand Convergence of Sensing, Communications, Computing, Storage, and Intelligence}
\author{\IEEEauthorblockN{Yuguang Fang,~\IEEEmembership{Fellow,~IEEE},  Yiqin Deng, ~\IEEEmembership{Member,~IEEE}, and Xianhao Chen, ~\IEEEmembership{Member,~IEEE}} 
\thanks{This work was supported in part by the Hong Kong SAR Government under the Global STEM Professorship and the Hong Kong Jockey Club under the Hong Kong JC STEM Lab of Smart City (Ref.: 2023-0108). The work of Y. Deng was supported in part by the National Natural Science Foundation of China under Grant No. 62301300. The work of X. Chen was supported in part by HKU-SCF FinTech Academy R\&D Funding. A complementary comprehensive version was accepted for publication~\cite{chen2024vehicle}.}.
\thanks{Yuguang Fang and Yiqin Deng are with the Department of Computer Science, City University of Hong Kong, Hong Kong, China (e-mail: \{yiqideng, my.fang\}@cityu.edu.hk).}
\thanks{Xianhao Chen is with the Department of Electrical and Electronic Engineering, the University of Hong Kong, Hong Kong, China (e-mail: xchen@eee.hku.hk).}
}


\maketitle

\begin{abstract}
The most commonly seen things on streets in any city are vehicles. 
However, most of them are used to transport people or goods. What if they also carry resources and capabilities for sensing, communications, computing, storage, and intelligence (SCCSI)? We will have a web of sensors to monitor the city, a network of powerful communicators to transport data around, a grid of computing power to conduct data analytics and machine learning (ML), a network of distributed storage to buffer/cache data/job for optimization, and a set of movable AI/ML toolboxes made available for specialized smart applications. This perspective article presents how to leverage SCCSI-empowered vehicles to design such a service network, simply called {\em SCCSI network}, to help build a smart city with a cost-effective and sustainable solution. It showcases how multi-dimensional technologies, namely, sensing, communications, computing, storage, and intelligence, converge to a unifying technology to solve grand challenges for resource demands from emerging large-scale applications. Thus, with SCCSI-empowered vehicles on the ground, over the air, and on the sea, SCCSI network can make resources and capabilities on the move, practically pushing SCCSI services to the edge! We hope this article serves as a spark to stimulate more disruptive thinking to address grand challenges of paramount importance. 
\end{abstract}


\section{Introduction}
Smart cities demand significantly powerful sensing, communications, computing, storage, and intelligence (SCCSI) capabilities everywhere. For example, modern cities are expected to install many video cameras for different purposes such as traffic monitoring and control, reckless driving detection, and security surveillance. Imagine that we leverage a set of surveillance cameras to help monitor potential crime scenes in a tough city neighborhood. Video clips captured from those cameras can be transported to the local police department in real-time and analyzed via data analytics, and then real-time decisions on criminal activities can be made, and control actions can be taken (e.g., dispatching police patrols to the scene). For this simple cyber-physical system to run effectively, we need sufficient spectrum resources from a crime scene to a police department, enough computing and storage space to analyze video streams and draw conclusive decisions, and enough personnel to take corresponding actions. As we observe, a lack of any resources (cameras, spectrum, computing, storage, and intelligence) will not make it possible to fulfill the whole mission of crime detection. This is the kind of problem we must address for smart city.


Smart city intends to provide the city with the following smart-* features: smart mobility, smart grid, smart health, smart living, smart environment, smart people, smart government, smart economy, etc., with the ultimate goal of improving people’s quality of life (QoL) \cite{leung2023implementing,kirimtat2020future}. The purpose is to provide its residents with more convenient daily routines, better lifestyles and living, improved healthcare and aging, and better inhabiting environments, which cannot be achieved by a single sector or a single city authority but by concerted and collective efforts of all participants. Tremendous studies on smart cities have been carried out in the last few decades, and many demonstrations have been conducted in many cities. However, many efforts have been mostly typified with single-dimensional efforts in the sense that they only show demonstrations of single-domain applications, such as smart mobility, city-wide pollution detection, smart energy, or road debris detection \cite{leung2023implementing,kirimtat2020future}. Besides, most proposed solutions have not considered cost. It takes a holistic approach to derive a global design of a smart city with long-term economic sustainability.


Unfortunately, accomplishing the global design is highly challenging, particularly with economically sound and sustainable solutions! In a nutshell, we have to handle various kinds of information collection and exchanges about the city. A large diverse set of sensing devices, such as smart meters, cameras, radars/lidars, air quality sensors, and even information content analyzers, should be deployed in the city to sense the ``pulse'' of the city. The resulting large volumes of data in the city may have to be collected and transported to certain locations where they can be consumed; various types of data may need to be temporarily stored or cached for further communications, processing, or computing for intelligence extraction; and powerful intelligence information has to be provided in a timely fashion to make city operations smart. All these salient services and capabilities are fundamental building blocks of a smart city, which cannot be achieved without a robust and concerted supporting framework for SCCSI. Particularly, communications capability should be provided to transport data around the city for better services (\eg, Internet services); computing capability should be enabled at the edge to take over the workloads from resource-limited customers' devices to reduce latency, lower energy consumption, and harvest in-situ intelligence for smart city operations (\eg, AR/VR navigation and autonomous driving); storage capability can be used to smooth out bursty service traffic and offer higher efficient services for both communications and computing, and data analytics provides intelligence needed for smart city operations and IoT applications. All these must work together interactively, collaboratively, and concertedly with the ultimate goal of guaranteeing the desired users' experience, which will not be possible without a holistic, systematic design. 

Although telecommunications industries (e.g., 5G/6G and beyond) and computing industries (e.g., cloud/edge computing) have all made significant strides to integrate information and communication technology (ICT) and computing technologies to deliver the promise of a smart city. Yet, they have not yielded long-term economically cost-effective and sustainable solutions. For a city to adopt a cellular solution such as Cellular Vehicle-to-Everything (C-V2X), the city has to recruit a cellular operator to build the infrastructure and run the cellular services for the smart city operations and services. This may require not only costly capital investment in infrastructure but also continuous spending on high operational and maintenance costs, which will not be cost-effective for a smart city of a reasonable size. We need to search for a cost-effective alternative. 

Our proposed design in this article is inspired by the following intriguing observations. The most popular things on the streets in any city are vehicles, which, if they choose, could reach every corner of the city. However, their mobility has not been fully utilized except for transporting goods or people. What if they carry devices with powerful SCCSI resources and capabilities? When they are roaming around a city, they could form a web of sensing, a mobile communication network, a grid of mobile computing servers, a distributed mobile storage network, and a system of AI services on the move. Such a device, we call {\em Point of Connection (PoC)} to provide SCCSI resources or capabilities, can be customized according to the need, massively produced just like WiFi routers, to lower the price, so that even the residents with reasonable income could afford. With PoCs populated in a city, installed on vehicles (public or private), at strategic positions (intersections or lampposts), collocated with base stations (BSs) for cellular systems or access points (APs) for WiFi \cite{chen2024vehicle}, over the air (on unmanned aerial vehicles or UAVs), high altitude platform (HAP) or aircraft, airships, even satellites) \cite{deng2024uav}, and on the sea (on vessels and ships), then we form a powerful edge network of PoCs with multi-dimensional resource capability, simply called {\em SCCSI network}, a special kind of space-air-ground-sea networks, to serve smart city operations and services (O\&Ss). Any resident could gain services from it for SCCSI capability and could offer SCCSI capability if he/she is driving a vehicle with PoCs. Since we proactively leverage crowdsourcing to build a SCCSI network for a smart city and the residents who contribute their SCCSI capability (the use of their PoCs) make contributions to the smart city O\&S, they could acquire city reward points for their contributions, cashable for their other city services (e.g., pay their utility bills). Thus, as we draw resources from people and then enable them to serve the people, the so-designed SCCSI network provides an economically cost-effective and sustainable approach to building a smart city. 

To achieve this goal, a sound and technically feasible information networking and computing framework with effective data, resource, and mobility management has to be carefully considered. Thus, \emph{a holistic investigation on various communications, networking, computing, and data management issues of smart cities has to be carefully conducted for its viability, usability, and cost-effectiveness in response to the envisioned design objectives}. In this article, we present how to leverage vehicles to develop such a novel disruptive service provisioning approach, called {\em Vehicle as a Service (VaaS)}, and deliver an economically sound and sustainable solution to enabling SCCSI services for the realization of future smart city. 

\section{Architectural Design of SCCSI Networks}
The key on the top is how to design the SCCSI network architecture to facilitate the envisioned operations and services. In this section, we present such an architecture, from the basic component design to the overall SCCSI network framework.  

\subsection{Point of Connection (PoC)}
\begin{figure}
\centering
\includegraphics[width=0.45\textwidth]{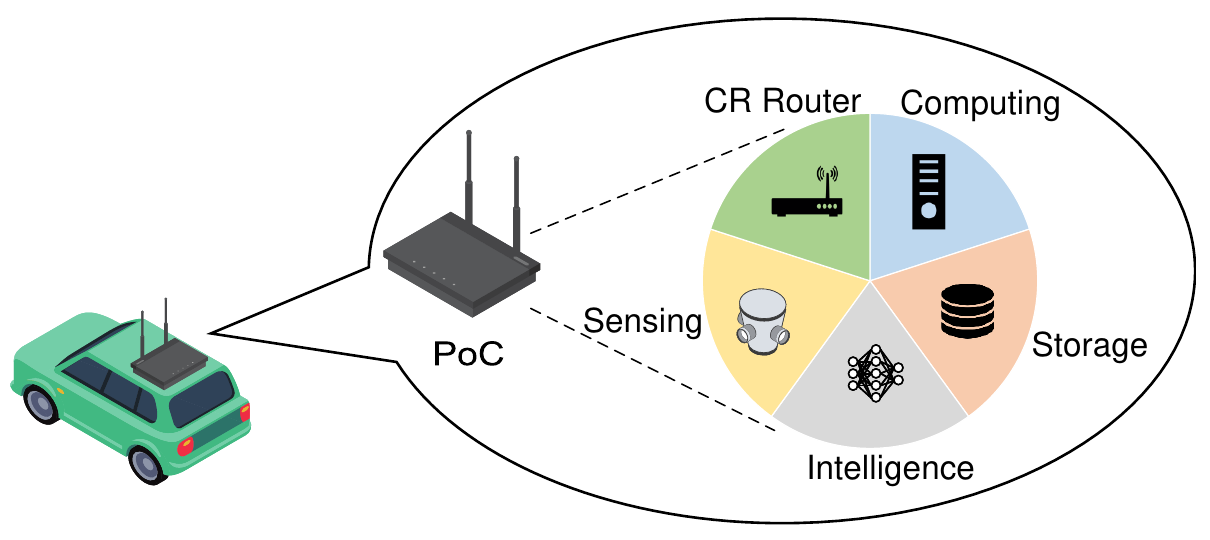}
    \caption{Logical design of PoC, a connecting device equipped with SCCSI capabilities.}\label{poc}
\vspace*{-0.1in}
\end{figure}

The fundamental building block is the introduction of the conceptual set-top box, namely, PoC, which is equipped with customized SCCSI resources or capabilities. As shown in Fig.\ref{poc}, the set-top box PoC consists of customized SCCSI resources and capabilities as described below. 

\begin{itemize}
\item{\em Sensing}: A PoC is equipped with interfaces with multi-modal sensors, which can be easily connected with multi-modal sensing capabilities within and outside vehicles, logically making a PoC have powerful sensing capabilities. 

\item{{\em Communications}}: A PoC is installed with a powerful cognitive radio router that can be used to harvest idle spectrum in its surrounding proximity for fast data communications. Since future smart city services involve real-time services with large data volume, fast data exchange of big data is unavoidable, hence powerful communications capabilities are in dire need. 

\item{{\em Computing}}: A PoC is equipped with a customized computing server, say, with GPUs if needed, to perform real-time data processing and computing or conduct data analytics. Vehicles carrying such PoCs form a computing server on the move, really pushing computing to the edge. 

\item{{\em Storage}}: A PoC is equipped with fast distributed networked storage for data storage, buffering, and prefetching/caching (e.g., as in information centric networking or ICN). Both communications and computing systems do need sufficient storage space to optimally carry out communications and computing tasks. 

\item{{\em Intelligence}}: Customized AI/Machine Learning (ML) toolboxes can be installed in a PoC. Specialized AI/ML modules for smart cities can be designed to facilitate citizens to run their modules for smart city services.   

\end{itemize}

\begin{figure}
\centering
\includegraphics[width=0.45\textwidth]{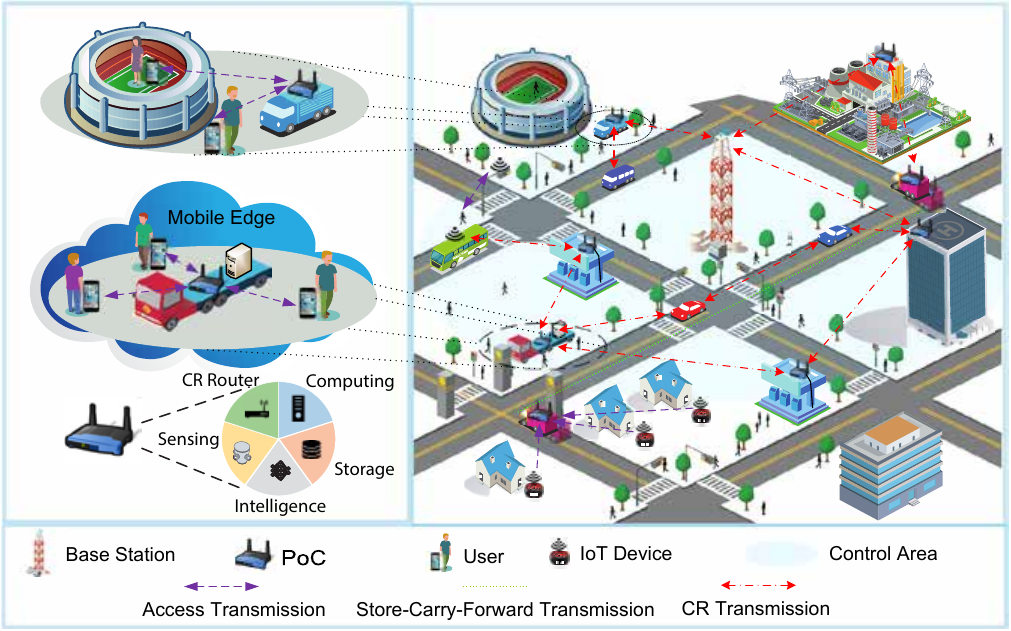}
    \caption{The overall system architecture of the SCCSI network.}\label{fig:system_model}
\vspace*{-0.1in}
\end{figure}

\subsection{SCCSI Network Formation}
Next, we articulate how to deploy PoCs to form our SCCSI network to support various smart city operations and services. As shown in Fig. ~\ref{fig:system_model}, we envision that we install PoCs at strategic locations, such as at roadsides or intersections, acting as roadside units (RSUs); on rooftops, acting as relays, or in vehicles on ground or over the air (UAVs), acting as mobile PoCs to store-carry-forward data. PoCs can also be installed at or co-located with (small) cellular BSs or APs. Depending on end customers' needs, PoCs can be customized accordingly. Fixed PoCs, such as RSUs, can be connected directly to the backbone  via wired or reliable wireless links of potentially high data rates. Local PoCs can form mesh networks to collectively handle data traffic. A PoC can be a traffic offloading point when fixed, or a data carrier when mobile. 

With such PoC deployment, the fixed PoCs on the backbone edge (PoCs at BSs/APs/RSUs) provide rich connectivity to the Internet and computing services. The mobile PoCs offer the backhaul mesh networks to connect end users to the backbone for various data services and computing services. The so-formed network of SCCSI capability-empowered PoCs, together with the backbone data networks, are generically called {\em SCCSI networks}. Moreover,  by leveraging both spectrum opportunity and mobility opportunity, \emph{our SCCSI network, particularly the mobile mesh of PoCs, could push SCCSI resources and capabilities much {\em closer} to the end users.} Thus, with proper deployment of PoCs, tremendous volume of data, the so-called {\it big data}, can be conveniently collected, processed, and/or transported around a city for timely decision making, which plays a crucial role in implementing the vision of IoT \cite{nguyen20216g,syed2021iot,atzori2010internet} and smart cities \cite{kirimtat2020future}. 

\subsection{Network Operator and Control Plane}
To run the SCCSI network, we introduce a new network operator, called {\em Secondary Service Provider (SSP)}, adopted from cognitive radio networks \cite{ding2017cognitive}. The reason we use ``secondary service'' is that a SCCSI network heavily relies on the resources harvested from other users or services (the primary services). SSP is responsible for PoC deployment along the backbone network edge, fulfilling the partial service coverage at strategic positions with fixed PoCs. It also runs the administrative management, operations, maintenance, accounting and billing, etc. It can be a cellular operator, if a city chooses a cellular operator to run this system, and it can be independent system integrator (SI) who helps the city implement and manage the SCCSI network for smart city operations and services. No matter what, it has to have fixed spectral resources (owned or leased) to support reliable control signaling to manage the network. Cellular operators, such as 6G mobile operators, can serve as SSPs when they wish to enhance their services by utilizing the ubiquitous resources from vehicles. Importantly, the SCCSI network is highly compatible with 6G in the sense that our SCCSI network is implemented with cognitive radio technologies, and its integrated AI and communication, as well as C-V2X technologies, will be inherently supported by 6G. This compatibility makes it easy for mobile operators to implement the framework proposed in this article.

\begin{figure}
\centering
\includegraphics[width=0.45\textwidth]{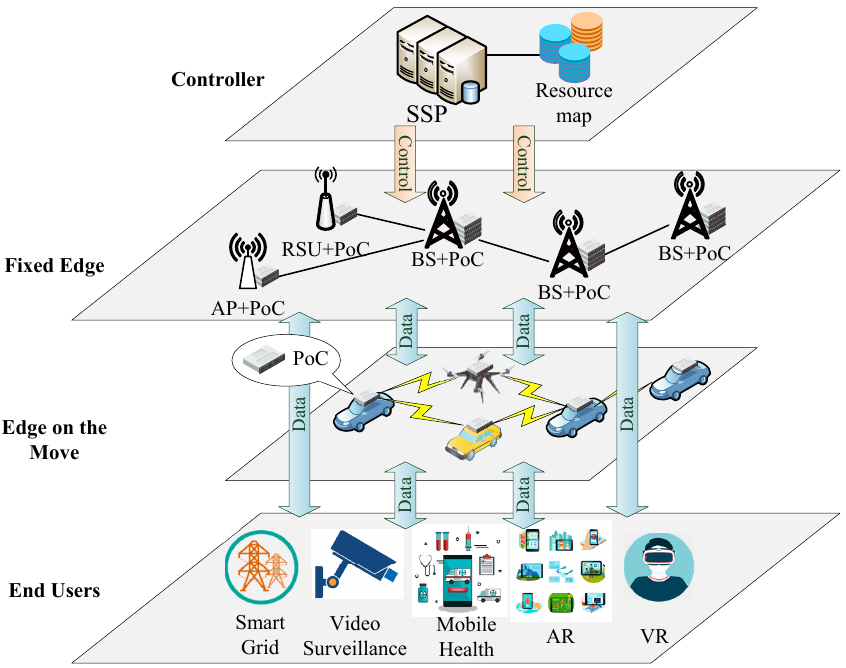}
    \caption{The overall system architecture of our SCCSI network. It attempts to integrate both fixed and mobile infrastructure to provide services. }\label{control}
\vspace*{-0.1in}
\end{figure}

Obviously, the effectiveness of the SCCSI network heavily relies on the availability of resource information in the network environment and opportunistic resources is only best used when resource state information can be quickly made available timely, which will not be possible in a fully distributed fashion. Thus, we do need a centralized control plane to manage the SCCSI network operations. In light of this observation, we will adopt the software-defined network (SDN) design methodology to design the control plane. Under this design, a selected subset of fixed PoCs are used to run the SDN software and proactively collect state information about SCCSI resources, requested traffic information, edge server load information, channel state information, etc., so that the state of the SCCSI network is fully known through the control plane. The overall logical design of our SCCSI is shown in Fig.\ref{control}. As we observe, we need to characterize various kinds of links with their functionalities carefully defined. 

\section{Incentive Design to Stimulate Participation}

SCCSI networks do need active participation of PoCs from all sources, particularly those from private vehicles' owners as the design philosophy is to acquire resources from the people and then make them available for the people.  Thus, it is imperative to design viable incentive to stimulate PoC owners to participate. The first issue is how to define the incentive value. It is hard to quantify how much it costs for a usage of a PoC (a resource or a service that a PoC offers) because it depends on many factors. For example, how much a PoC is rewarded when it relays a data item? It may depend on the type of data item being delivered. For urgent message, the cost may be higher while for best effort message, it may cost minimal. Besides, it also depends on timeliness and PoC population around the helping spot. Thus, it is even harder to quantify the value at decision time. One possible way to do is to leverage the reward point system, like flight mileages, hotel points, or grocery shopping points, in which point value is uncertain and can be determined and/or redeemed in the future. We propose to use such point rewarding system in our SCCSI network. 

For a smart city, a reward point system is the perfect fit \cite{yi2003effects}. The SCCSI network is to provide smart city operations and services, thus whatever help a PoC provides to SCCSI services is to help city operations, and thus the point earned can not only be redeemable for SCCSI services but also is used to pay other utility bills! 

With incentive valuation fixed, the next issue is to address how to enable people to help each other for SCCSI service provisioning. One option is to let SSP handle the issue in a centralized fashion: determine what resources are needed at certain spot at certain time and search for sufficient resources to meet the demanded service requirements. Another option is to let service provisioning run in a distributed fashion: a SCCSI service requester (buyers) determines how much resources are needed and makes offer with points to pay, while PoC owners who have the requested resources (sellers) responds with bids. Clearly, either case can be handled with auction theory. 
Obviously, we need to design viable auction mechanisms to run the SCCSI network for a smart city. 

\section{How to Proactively Leverage the SCCSI Resources and Capabilities}
With sufficient participation of SCCSI-empowered vehicles, the network segments formed by the vehicles carry SCCSI resources via PoCs roaming around the city, providing a rich set of resources, and, more importantly, pushing the resources and capabilities to the edge. The problem now is how to take the opportunities this network offers and make the best use of them. In this section, we will present a few important opportunities to leverage such resources. 

\subsection{Leverage Spectrum Opportunity}
The reason we elect to use cognitive radio routers in PoC is that PoCs can serve as the agents for end users to deliver and receive data using harvested spectrums \cite{ding2017cognitive}. It is expected that smart city operations and services will generate a huge amount of data to be carried. No matter how much spectrums allocated to the city via cellular systems such as 5G/6G and beyond, they will not be enough as the exponentially growing mobile traffic is still expected. Besides, although such traffic is high, its traffic composition has salient features: a large chunk of Internet traffic such as video clips is delay-tolerant! For example, video clips occupies over $70\%$. If we can shift delay-tolerant traffic to the unreliable harvested spectrums while saving reliable spectrums for time-sensitive traffic, we can better serve residents with better quality of experience (QoE). More discussions can be found in \cite{ding2017cognitive}. 

As we alluded in \cite{ding2018smart}, PoCs on vehicles are more powerful than typical end users, we could leverage vehicular PoCs to conduct collaborative spectrum sensing. With SCCSI-empowered vehicles, city spectrum maps can be easily collected just as the city traffic map updates. Better yet, we could construct a spatio-temporal SCCSI resource map for the city so that service provisioning can be done quicker and better. With this in mind, a digital twin model can be constructed to serve smart city operations better. 

\subsection{Leverage Vehicular Mobility Opportunities}
Since many SCCSI network nodes are mobile, we can take advantage of vehicular mobility to provide SCCSI resources \cite{deng2024uav}. 

PoCs are designed to have the ready interfaces of multi-modality sensing capabilities, thus can easily access to sensors in vehicles. Thus, each vehicle with PoC, particularly an autonomous vehicle, can gather tremendous sensing data to share with others. With such rich data sources, powerful data analytics can be applied to improve driving safety, congestion control, and carbon reduction. With cameras installed in vehicles, SCCSI-empowered vehicles can collaboratively detect events on streets (e.g., crime in progress, person in heart attack, etc.) and take actions without physically intervening. For example, pure presence of multiple vehicles may deter crime in progress. 

For data transportation, delay-tolerant traffic can be offloaded to nearby vehicular PoCs, carried over to a location with higher network capacity for offloading. Data can be moved to the curb of the regular routes for scheduled public transits to pick up just as to pick up passengers. One interesting problem is how to efficiently transport data over the city to meet smart city demands. A closely related subproblem is how to cache contents and how to manage the information cache to minimize the city network traffic.  With the emergence of shared mobility, future autonomous vehicles may be used to fulfill the last mile customer delivery, in the sense that public transit will pick up the last mile services with the shared autonomous vehicles to realize the door-to-door services via such a public transit system. 

SCCSI network can facilitate computing. As PoCs are on the move, hence are the SCCSI resources. When a spot demands computing power, vehicles afar can be incentivized to either move to the spot to add more computing power, or vehicles close by can carry or relay computing load to places where computing power is high. For instance, to shape the traffic to avoid or control congestion, SCCSI-empowered vehicles can be incentivized to form platoons to increase computing power to perform data analytics and derive effective control strategies to improve driving safety and experiences. In a sense, vehicles with PoCs can be incentivized to gather together to boost computing power and be dispatched to locations where computing power is needed. One of the more interesting use cases of such a kind is to leverage UAVs to relay computing tasks over the air to reach afar vehicles~\cite{deng2024uav,deng2024uav-assisted}. In smart cities, there may be discrepancies between the distributions of SCCSI-empowered vehicles (i.e., resource supplies) and resource demands. By establishing line-of-sight channels at elevated altitudes, UAVs can serve as relays to alleviate such inconsistency by relaying/reflecting bursty computing tasks to the areas where  more SCCSI-empowered vehicles can help.

For rich distributed storage provided by our SCCSI network, we can form a naturally formed information centric network \cite{yue2014dataclouds}. Fixed PoCs form the information cache buffers while mobile PoCs can be used to carry information content to update at leisure (still timely with high probability). Whenever buffering spaces for either communications or computing are needed at certain spot, the surrounding vehicles with PoCs can be incentivized to move to the spot to help. This applies to the AI/ML toolboxes as well. With plenty of AI/ML toolboxes populated via PoCs, one can access the toolboxes and extract useful information about the smart city. 

It is noted that many smart city applications, such as connected and autonomous driving, high-definition (HD) map update/construction, and public safety, may demand multiple SCCSI resources to support their missions, as elaborated in \cite{chen2024vehicle}. It will be highly interesting to find the right combinations of PoCs to meet the mission requirements. 

\subsection{Leverage Opportunistic Capability In Situ and In Tempore}
With highly dynamic movement of mobile vehicles, the SCCSI network is also highly dynamic. There may emerge some network formations offering powerful SCCSI resources and capabilities. For example, roadside parked vehicles, if battery energy can be controlled to power PoCs,  provide stable clusters of computing servers and storage networks. SCCSI-empowered vehicles and autonomous vehicles in parking lots hold tremendous computing power and storage, which can serve as temporary computing centers or data centers to serve others. A platoon of autonomous vehicles will not consume much computing power when platooning, and thus, the saved computing power can be used as platooning cloudlets to perform computing tasks for others. Autonomous vehicles caught in traffic congestion may use minimal computing power, and the remaining powerful computing capacity in autonomous vehicles caught in traffic will be wasted if not untapped for good use for others. Particularly, with a proper incentive mechanism running in the SCCSI network, vehicles tend to flock more regularly. How to leverage such emergent in situ and in tempore SCCSI resources to provide smart city operations and services will be a challenging yet fruitful research topic. 

\section{Research Challenges and Future Research Directions}

To implement the envisioned SCCSI network for a smart city, we face many design challenges, leading to a few fruitful research directions. 

\subsection{Security and Privacy}
The first on the top is security and privacy. Since the fundamental building block is mostly privately owned vehicles and the design approach is to enable people to help people, all SCCSI service provisioning has to outsource others. For example, communications and computing may have to be carried out by others, and thus it is hard to protect the data to be communicated and computed. Privacy poses more serious concerns for this network. Data captured by the sensing devices may contain others' private data. For example, one's camera may capture people's private moments that may become viral if not properly managed, which may cause the city potential law suits. How to effectively handle the security and privacy over our SCCSI network is still wide open. Fortunately, SCCSI network participants are all registered users with their vehicles tagged, which is significantly different from other IoT applications. PoCs can work together to form the first line of defense and jointly manage security and privacy, say, running advanced AI algorithms to detect malicious attacks. To provide location privacy, a major concern for most vehicle owners, we can follow the privacy-preservation design in \cite{sun2010identity} to blind a vehicle's ID unless misbehaving, achieving privacy desired by vehicle owners and traceability required by law enforcement authorities. A good research direction is to leverage salient features of SCCSI networks to design effective strategies to protect security and privacy.   

\subsection{Incentive Mechanism Design}
Even if we remove the stumbling blocks of security and privacy, we still face the participation problem with privately owned vehicles. As we mentioned, we could design point reward systems and auction mechanisms to stimulate vehicle owners to contribute their resources. To attract long-term participation of vehicle owners, a point-based loyalty program can be employed to capture the psychological effects of participants in a point reward system. In this respect, we need an effective point system and potentially a new auction theory, which forms another rich research direction. 

\subsection{Mobility Management}
Due to vehicular mobility, mobile PoCs will cause service interruption. For example, if a mobile PoC is serving as a relay, moving away from the receiving end, the link quality will become weaker and weaker, so a handover has to be made just as in cellular systems. If a mobile PoC is used for a computing task, working together with another group of PoCs, when it is moving away from the group, its computing load has to be handed over to another node (computing service handover). Sometimes, multi-hop task routing \cite{deng2023multi} is necessary to maintain the end-to-end computing link, which is highly challenging. Since services delivered over a SCCSI network tend to be multi-dimensional in the sense that multiple resources may be simultaneously utilized, the mobility management in SCCSI networks is much more complicated and demands a careful investigation.  

\subsection{Emergent SCCSI Resource Clustering Dynamics}
As we alluded earlier, due to the controlled/incentivized mobility and incentive mechanism, SCCSI network nodes tend to have emergent clustering of PoCs. Such dynamics depends not only on vehicular mobility, but also on service demands together with incentive design. It would be very useful to create such resource clustering maps that such an in situ and in tempore SCCSI resources can be best utilized timely. 

\subsection{Traffic Shaping and Congestion Control}
It is well-known that the existence of a certain percentage of autonomous vehicles can help reshape the traffic flow \cite{wu2018stabilizing}. It is well-known that congestion tends to be caused by bursty traffic. With our SCCSI network, participating vehicles can be controlled or incentivized to execute certain control policies to regulate traffic flows in a smart city. Since incentivized SCCSI-empowered vehicles can be recruited and commended to serve as distributed controllers in a traffic flow, the burstiness of a traffic flow can be regulated to avoid traffic jams. However, how to design such control algorithms to overcome city congestion will be a challenging but important research problem. 

\section{Conclusion}

With innovative digitalization and AI empowerment, vehicles can be leveraged to carry powerful resources and capabilities to support integrated sensing, communications, computing, storage, and intelligence (SCCSI) for building a smart city. Therefore, this article has advocated Vehicle as a Service (VaaS) design that leverages vehicles' empowered capabilities to build up such a service network for smart cities. We hope this perspective article can inspire stakeholders to work together on VaaS design and develop an economically sound and sustainable approach to building smart cities. 

\renewcommand\refname{References}
\bibliographystyle{IEEEtran}
\bibliography{IEEEabrv,fang,vnet}

\begin{IEEEbiographynophoto}{Yuguang Fang}(Fellow, IEEE)
received an MS degree from Qufu Normal University, China, in 1987, a PhD degree from Case Western Reserve University in 1994, and a PhD degree from Boston University in 1997. He joined Department of Electrical and Computer Engineering at University of Florida in 2000 as an assistant professor, then was promoted to associate professor in 2003, full professor in 2005, and distinguished professor in 2019, respectively. Since 2022, he has been the Chair Professor with Department of Computer Science at City University of Hong Kong. 
He is a fellow of ACM and AAAS.

\begin{IEEEbiographynophoto}{Yiqin Deng}(Member, IEEE) received 
an MS degree in software engineering and a PhD degree in computer science and technology from Central South University, Changsha, China, in 2017 and 2022, respectively. She was a visiting researcher at University of Florida from 2019 to 2021. She was a postdoctoral researcher with School of Control Science and Engineering at Shandong University from 2022 to 2024. She is currently a postdoctoral research fellow with Department of Computer Science at City University of Hong Kong.  
\end{IEEEbiographynophoto}

\begin{IEEEbiographynophoto}{Xianhao Chen}(Member, IEEE) received the BE degree in electronic information from Southwest Jiaotong University in 2017, and a PhD degree in electrical and computer engineering from the University of Florida in 2022. He is currently an assistant professor with the Department of Electrical and Electronic Engineering, the University of Hong Kong. 
\end{IEEEbiographynophoto}

\end{IEEEbiographynophoto}

\end{document}